\documentclass[a4paper,11pt]{article}
\usepackage{pos}

\usepackage{natbib}
\title{Five Years of Mini-EUSO Observations from the ISS: Summary of Key Results}
\author*[a]{Matteo Battisti}
\onbehalf{for the JEM-EUSO Collaboration\\[-1mm]{\normalsize \normalfont (a complete list of authors can be found at the end of the proceedings)}}

\affiliation[a]{Istituto Nazionale di Fisica Nucleare (INFN), Sezione di Roma Tor Vergata, Rome, Italy\\
}
\emailAdd{mbattist@roma2.infn.it}
\abstract{Mini-EUSO is the first space-borne detector of the JEM-EUSO (Joint Exploratory Missions for Extreme Universe Space Observatory) program, operating on the International Space Station (ISS) since October 2019. Designed to search for Ultra-High Energy Cosmic Rays (UHECRs) above 10$^{21}~$eV and capable of placing a stringent upper limit on their flux at these extreme energies, paving the way to future space-based UHECR observatories, Mini-EUSO has completed more than 150 observation sessions over five years, accumulating approximately 750 hours of data. The mission has produced the first global UV emission maps of Earth and provided valuable insights into lightning phenomena and Transient Luminous Events (TLEs), such as elves, as well as artificial light sources and meteors. Notably, Mini-EUSO has conducted the first systematic space-based meteor survey, detecting over 22,000 meteors and identifying three interstellar candidates.
Among the observed TLEs, the most interesting class of phenomena are elves, which appear as expanding ring-shaped structures occurring at $\sim$90 km altitude. Mini-EUSO has detected elves with varying structures and different numbers of concentric rings, from single-ring up to five rings. Thanks to its imaging capabilities, fast time resolution, and favorable observational geometry, Mini-EUSO is uniquely suited to studying this kind of lightning phenomena, providing unprecedented insight into their dynamics.
Additionally, the instrument has demonstrated the capability of a space-based detector to identify short light transients resembling extensive air shower signals while distinguishing them from those produced by UHECRs.
This contribution presents a comprehensive summary of the Mini-EUSO mission, its status, and main results.}

\ConferenceLogo{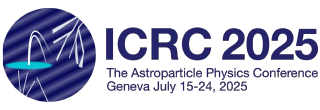}

\FullConference{39th International Cosmic Ray Conference (ICRC2025)\\
 15–24 July 2025\\
Geneva, Switzerland\\}
\begin{document}
\maketitle

\section{The detector}
Very large fluorescence telescopes equipped with highly dynamic electronics can detect a wide variety of phenomena across different classes and physical domains, significantly extending the scientific reach of these experiments well beyond the cosmic rays physics. In particular, they can contribute to atmospheric science through the observation of various lightning phenomena, to planetary science via meteor detection, and to fundamental physics through searches for exotic phenomena such as macroscopic dark matter, among other exploratory objectives. Mini-EUSO
offers the opportunity to validate the broad range of the scientific potential of a space-based observatory for UHECRs.

\begin{figure}[hbtp]
\centering
\includegraphics[width=.98\textwidth]{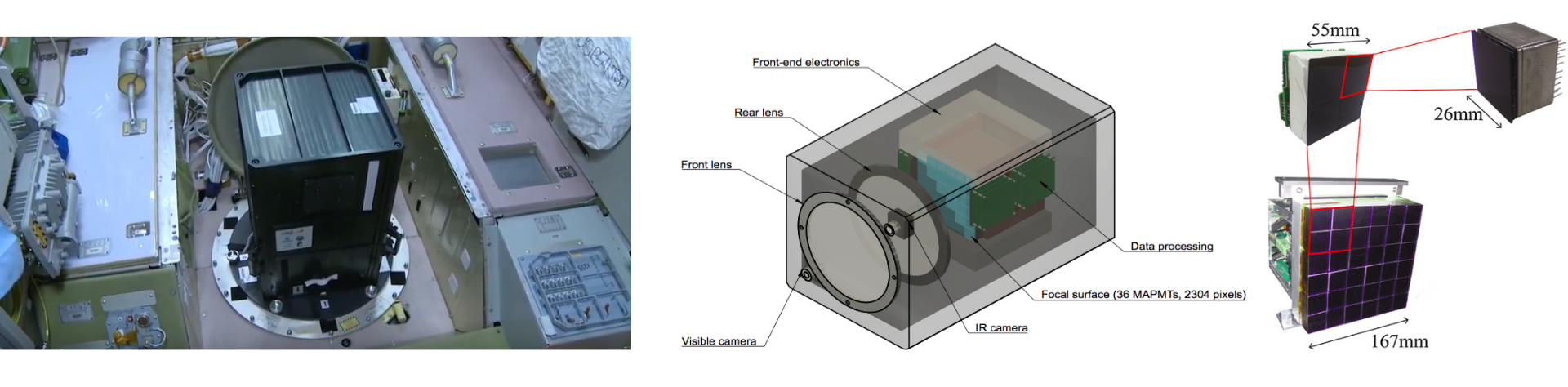} 
\caption{\textbf{Left:} Mini-EUSO installed on the UV-transparent window during a data-taking session. \textbf{Middle:} Schematic view of the instrument. \textbf{Right:} The focal plane, made of an array of MAPMTs, for a total of 2304 pixels arranged in a $48\times48$ matrix.}
\label{fig:Mini-EUSO_FlightModel}
\end{figure}

 Mini-EUSO is the first detector of the JEM-EUSO program \cite{JEM-EUSO_Zbigniew} operating in space. Mounted on the Zvezda module of the ISS, Mini-EUSO observes the Earth through a nadir-facing UV-transparent window. The size of this window determines the optical system, which consists of two 25~cm diameter Fresnel lenses. The main camera features an array of Multi-Anode Photomultiplier Tubes (MAPMTs) arranged in a 48$\times$48 pixel matrix (Fig.~\ref{fig:Mini-EUSO_FlightModel}), mostly sensitive to the near-ultraviolet (290-430 nm band) working in single photon-counting mode. 
 %The instrument’s 44$^\circ~\times~$44$^\circ$ field of view corresponds to a 300~km$~\times~$300~km area on the ground, with a pixel resolution of $\sim$6~km.
Mini-EUSO has been designed to detect a photon count rate per pixel per unit time from diffuse sources (nightglow, clouds, cities, etc..) similar to that expected from a large mission in space such as POEMMA~\cite{poemma} or the original JEM-EUSO \cite{JEM-EUSO_telescope} mission. The pixel footprint is $\sim$100 times larger than the original JEM-EUSO detector,  resulting in $\sim$6$\times$6~km$^2$ on ground, to compensate for the smaller optical system constrained by the dimension of the UV transparent window (covering an overall area of $\sim 350 \times350~$km$^2$ on ground). The consequence of the relatively small optical system and the large field of view of each pixel is that the energy threshold for the detection of UHECRs is above $10^{21}$~eV~\cite{Mario_exposure_EPJ}. 
Mini-EUSO unique acquisition system allows it to observe the Earth simultaneously on three different timescales, with time resolutions of 2.5~$\mu$s (called D1), 320~$\mu$s (called D2), and 40.96~ms (called D3). D1 and D2 are triggered, while in D3 Mini-EUSO takes data continuously.

The first observations took place on October 7, 2019. Since then, more than 150 sessions have been performed. Data are stored locally on USB pen-drives. After each data-taking session, samples of data (about 10\% of stored data) are copied and transmitted to ground to verify the correct functioning of the instrument and subsequently optimize its working parameters, while the USB drives containing all stored data are then returned to Earth on the first occasion. Currently, more than 750 hours of data are available on the ground from 5 years of operation.

\section{End-to-end calibration}
\begin{figure}[hbtp]
\centering
\includegraphics[width=.35\textwidth]{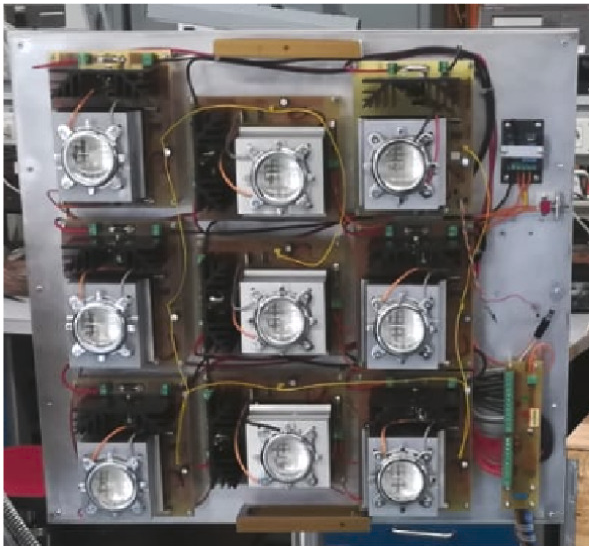} 
\includegraphics[width=.43\textwidth]{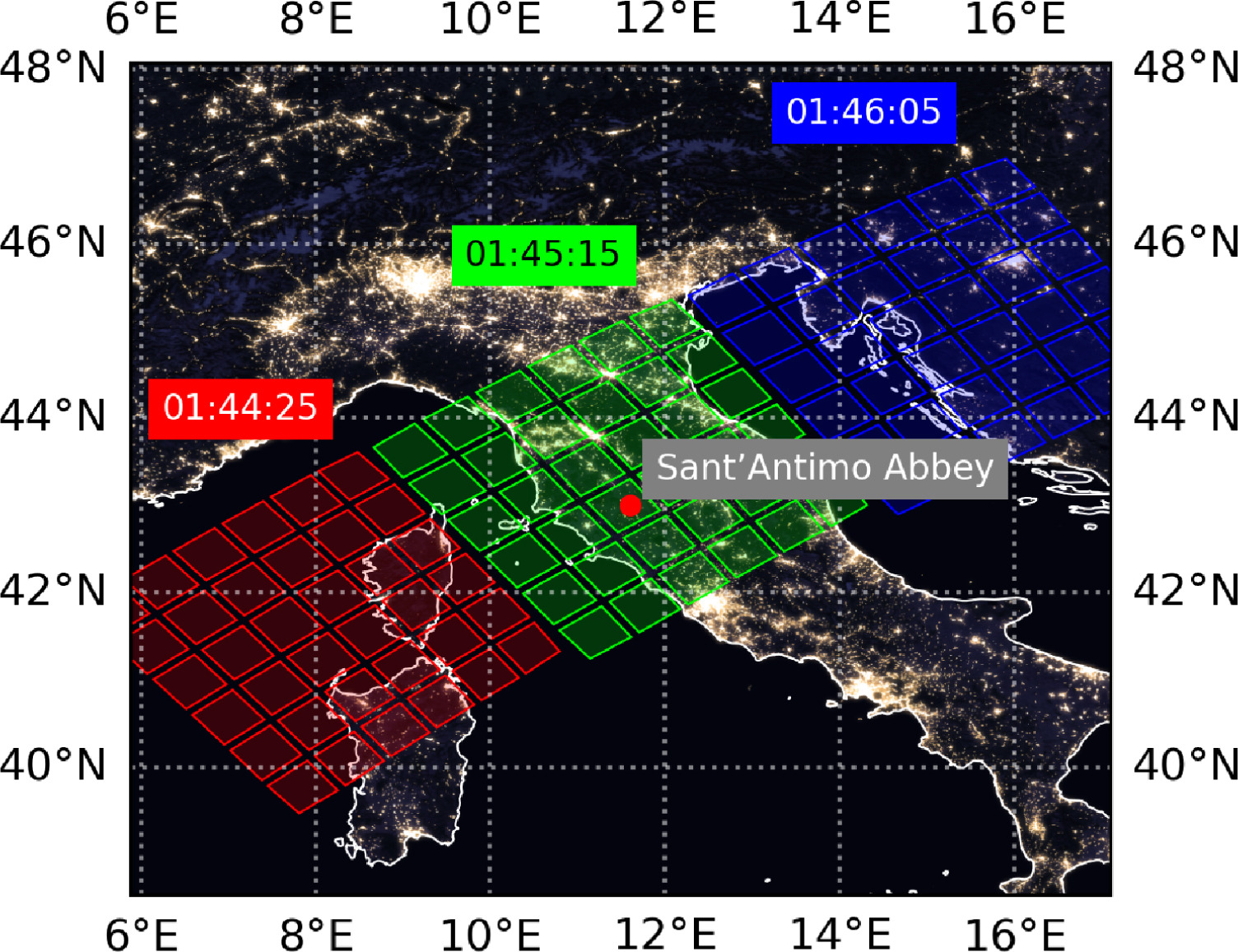} 
\caption{\textbf{Left:} Picture of the flasher used for the end-to-end calibration. \textbf{Right:} Foreseen FoV of the Mini-EUSO telescope superimposed to the geographical map for the passage over Sant’Antimo Abbey, on October 30, 2022. The background picture is taken from the Black Marble 2016 dataset available at \url{https://visibleearth.nasa.gov/images/144898/earth-at-night-black-marble-2016-color-map}}
\label{fig:UV_flasher}
\end{figure}

One of the main goals of Mini-EUSO is to measure the UV emissions from the ground and atmosphere and to assess the feasibility and performance of the measurement of UHECRs by means of a space-based detector. That requires the knowledge of the correlation between the observed photon count rate and the number of photons received, namely the absolute calibration of the instrument. During the past years, a few observational campaigns have been completed, employing a ground-based UV flasher to perform an end-to-end calibration of the instrument.  A successful campaign took place on October 30, 2022, when the flasher was located next to the Sant'Antimo Abbey, in central Italy, and observed by Mini-EUSO (Fig.~\ref{fig:UV_flasher}).

\begin{figure}[hbtp]
\centering
\includegraphics[width=.8\textwidth]{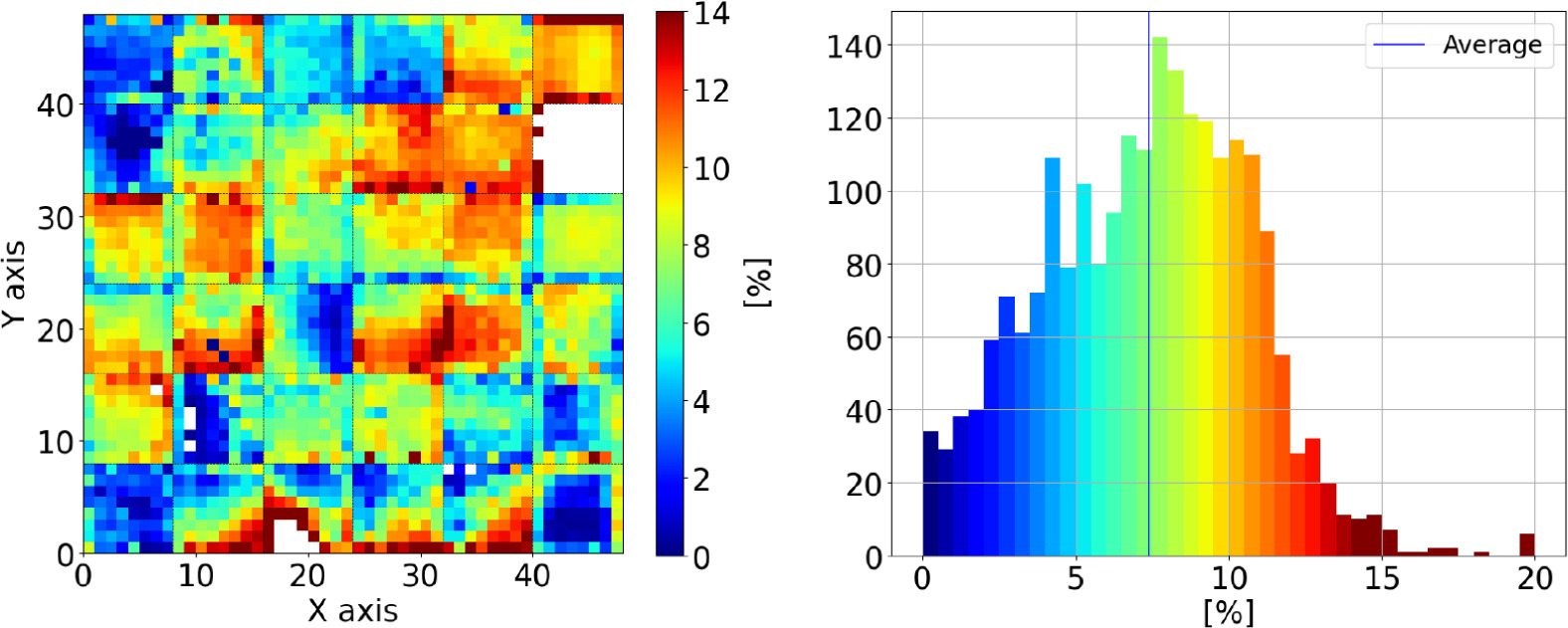} 
\caption{ Efficiency of the Mini-EUSO focal surface. \textbf{Left:} Map of the efficiency. \textbf{Right:} histogram of the efficiency values. The color scale (in percent) is common to the two panels. The average value is 7.3\%. More than 50\% of the pixels have an efficiency higher than 6\% and lower than 11\%.}
\label{fig:Efficiency}
\end{figure}

The flasher was previously calibrated in the lab, obtaining the number of emitted photons at a given wavelength (400 nm) per unit of time. Comparing the number of emitted photons with the number of detected photolectrons (and taking into account the transmission factors in the atmosphere for the specific weather conditions of that night) it is possible to obtain the absolute calibration value of all the pixels that observed the flasher. From there, the efficiency for all the other pixels can be derived (Fig.~\ref{fig:Efficiency}). A complete and detailed analysis of the calibration procedure can be found in \cite{Calibration_paper}.

\section{UV maps and exposure}
\begin{figure}[hbtp]
\centering
\includegraphics[width=.65\textwidth]{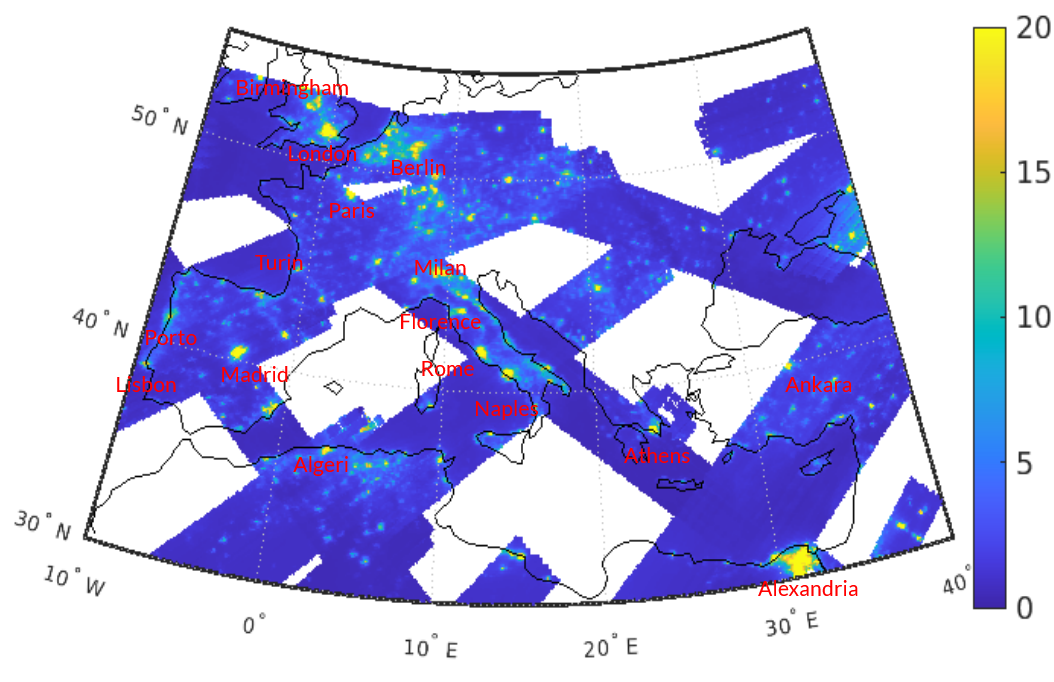} 
\caption{Counts observed in moonless conditions over Europe and the Mediterranean area. Sparsely populated areas, like the Sahara desert and the Carpathian and Apennine mountains, appear darker than the surroundings.}
\label{fig:UV_map}
\end{figure}

Correlating the counts observed by the detector with the location on ground, it is possible to produce %high-resolution
maps of the UV emission of the Earth with a sub-pixel spatial resolution. A map of the Mediterranean area in moonless condition is shown in Fig.~\ref{fig:UV_map}. 
The database of the UV-maps is publicly available at \cite{UV_dataset}, while the details on the creation of the maps as well as the correlation of the observed counts with environmental condition is detailed in \cite{UV_maps}.

\begin{figure}[hbtp]
\centering
\includegraphics[width=.95\textwidth]{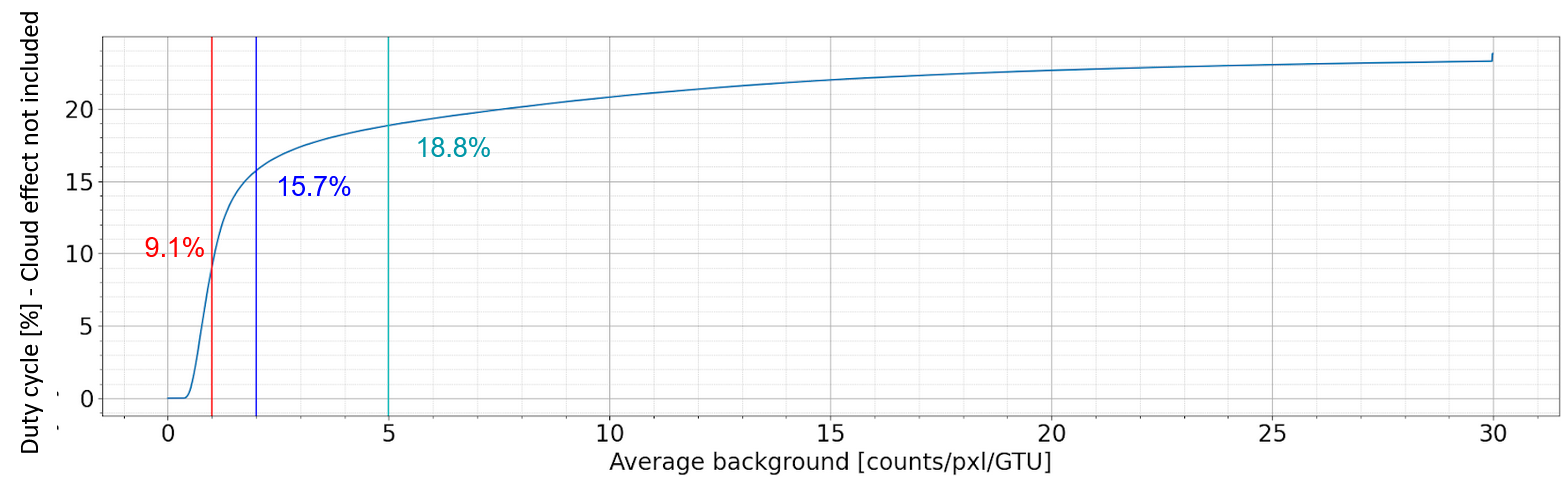} 
\caption{Percentage of time that a space-based UHECR detector will spend below a given background rate, taking into account the day and Moon cycle, as well as the presence of cities and other anthropogenic sources.}
\label{fig:Duty_cycle}
\end{figure}

Going further in the analysis, it is possible to use Mini-EUSO data to estimate the expected background for a future, larger, space-based UHECR mission such as the original JEM-EUSO telescope, the POEMMA detectors or the newly proposed M-EUSO mission. %The first step is to use Mini-EUSO data to estimate the background observed by these detector
The result is shown in Fig.~\ref{fig:Duty_cycle}.
 The plot highlights that a space-based detector will observe for the $\sim$9\% ($\sim$16\%) [$\sim$19\%] of the total time a background below 1 (2) [5] counts/pixel/GTU. These numbers take into account the day/night cycle, the presence of the Moon, and anthropogenic sources. It does not take into account the presence of the clouds. More details on this aspect can be found in \cite{Exposure_Mario}.

\section{Meteors}
\begin{figure}[hbtp]
\centering
\includegraphics[width=.43\textwidth]{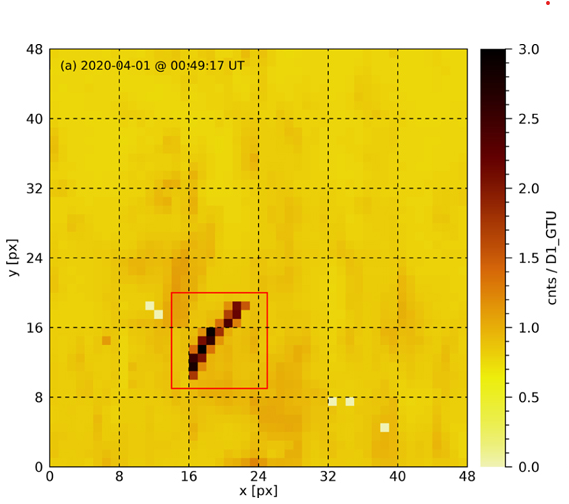} 
\includegraphics[width=.56\textwidth]{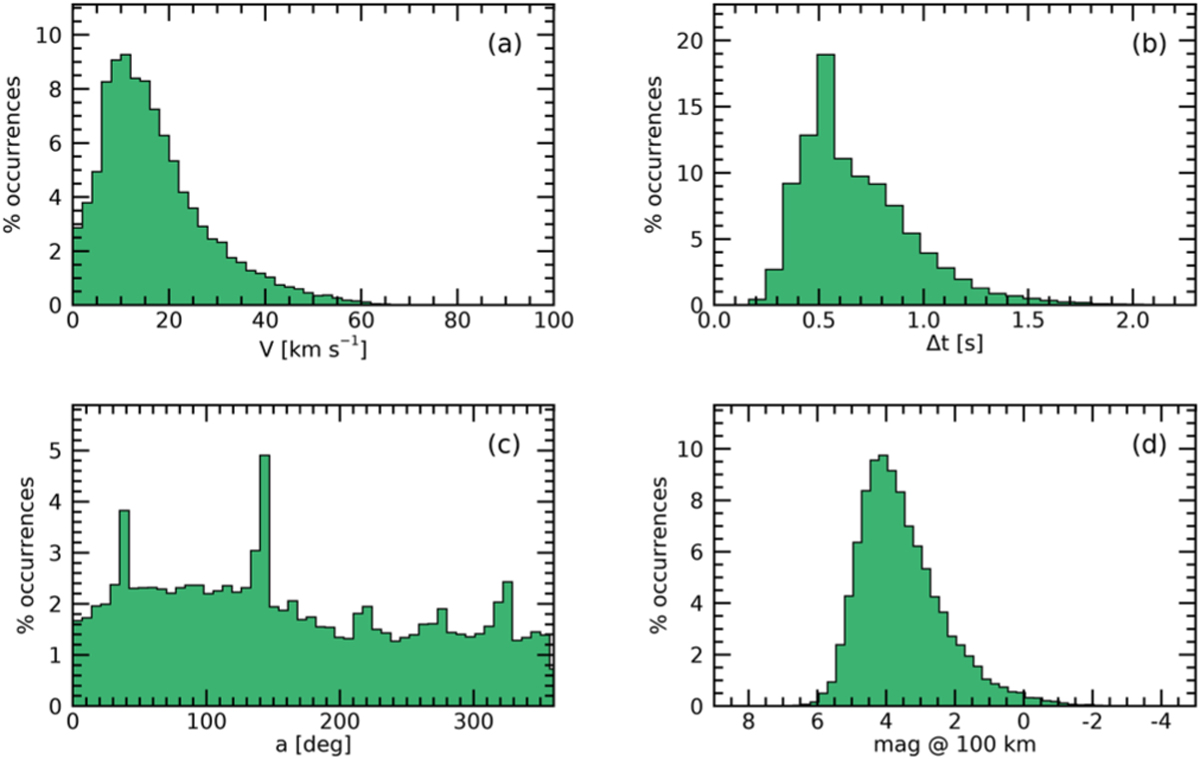} 
\caption{\textbf{Left:} Integrated track of a meteor. The event lasted for 32 D3 GTUs ($\sim$1.3 s). \textbf{Right:} Distribution of the physical parameters of 24 thousand meteors. (a) Horizontal speed; (b) duration of the event; (c) arrival azimuth angle; and (d) minimum absolute magnitude.}
\label{fig:Meteor}
\end{figure}

Mini-EUSO has performed the first ever systematic observation of meteors from space.
Meteors appear in Mini-EUSO data as point sources moving at few tens of km per second and lasting for hundreds of ms, up to few seconds. They are therefore observed in D3 dataset, with a time resolution of 40.96 ms. An example is shown in Fig~\ref{fig:Meteor}.
In the dataset analyzed so far, corresponding to less than 50\% of the total Mini-EUSO data, over 24,000 meteor events have been found, at an average rate of 2.92 per minute. If Mini-EUSO had taken data continuously over the entire 5-years period, we could expect over 2 million meteors, taking into account the detector duty cycle and the presence of clouds and moon light. As shown in the figure, Mini-EUSO starts detecting meteors at magnitude +6. Despite no events have been found with horizontal speed above 71 km/s, there are 3 events that are candidates for interstellar meteors, on which the analysis is still ongoing. A detailed report on the analysis of the meteor database is available in \cite{meteor_Mini-EUSO}.

\section{Elves}
\begin{figure}[htbp]
\centering
\includegraphics[width=.38\textwidth]{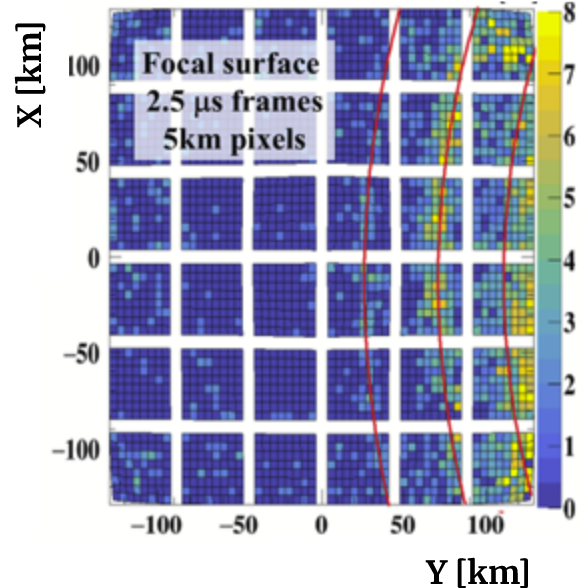}
\includegraphics[width=.61\textwidth]{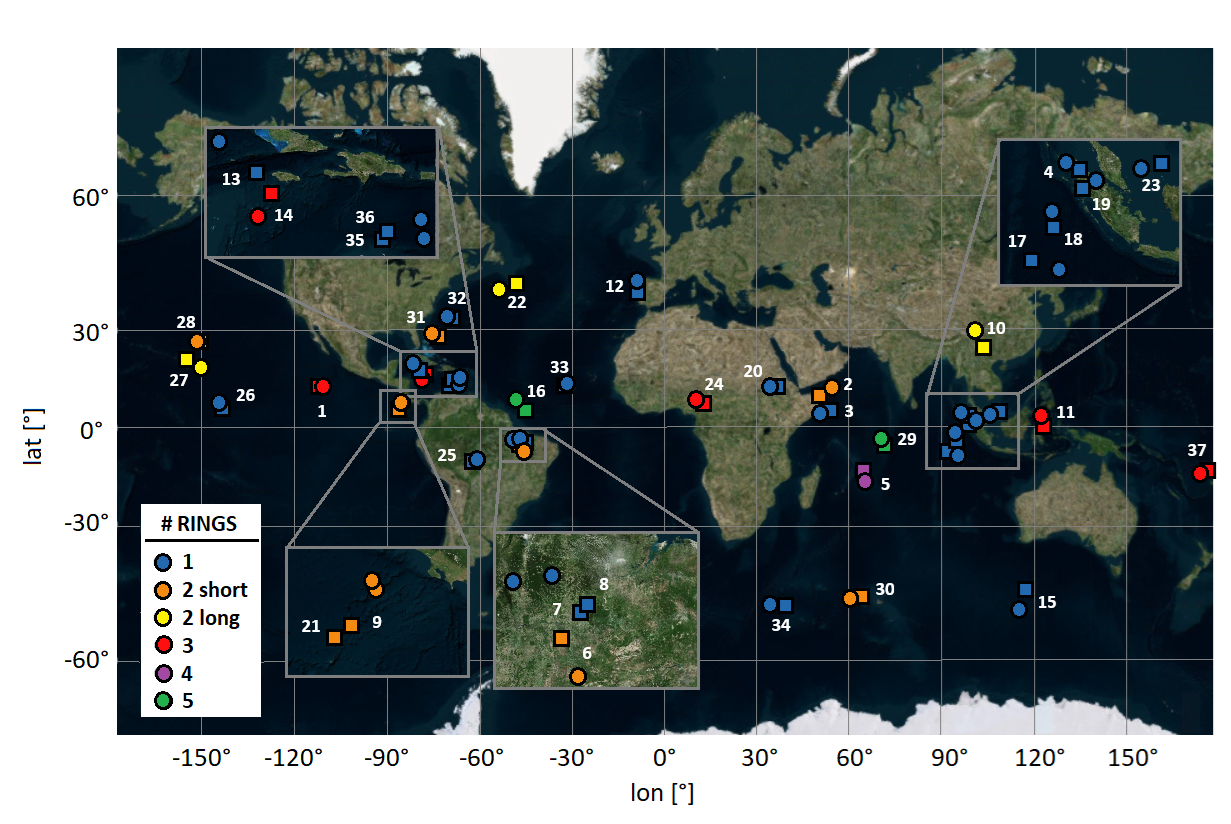}
\caption{\textbf{Left:} A 3-ringed elves detected by Mini-EUSO. The view shows the projection of the detected counts at 90~km altitude. \textbf{Right:} Map of the detected elves. The colours denote the number of rings for each event. Both the position of the detector (square marker) and the reconstructed position of the centre of the elves rings (circle marker) are shown.  Most of the events  are distributed in  the equatorial region.}
\label{fig:ELVE-map}        
\end{figure}

Elves appear in Mini-EUSO as fast expanding rings, propagating over hundreds of km and lasting for hundreds of $\mu$s (Fig.~\ref{fig:ELVE-map}, left). Thanks to its excellent time resolution, Mini-EUSO can follow their evolution as they propagate in the ionosphere. Min-EUSO detected 37 elves  (Fig.~\ref{fig:ELVE-map}, right) in 160 hours of data acquisition performed between November 2019 and October 2022, identifying events with radial extension up to $\simeq 800$~km and made of up to 5 concentric rings. %More details on this topic are available in \cite{Elves_Laura_Matteo}.

\section{Short Light Transients (SLTs)}
\begin{figure}[h!]
\centering
\includegraphics[width=.99\textwidth]{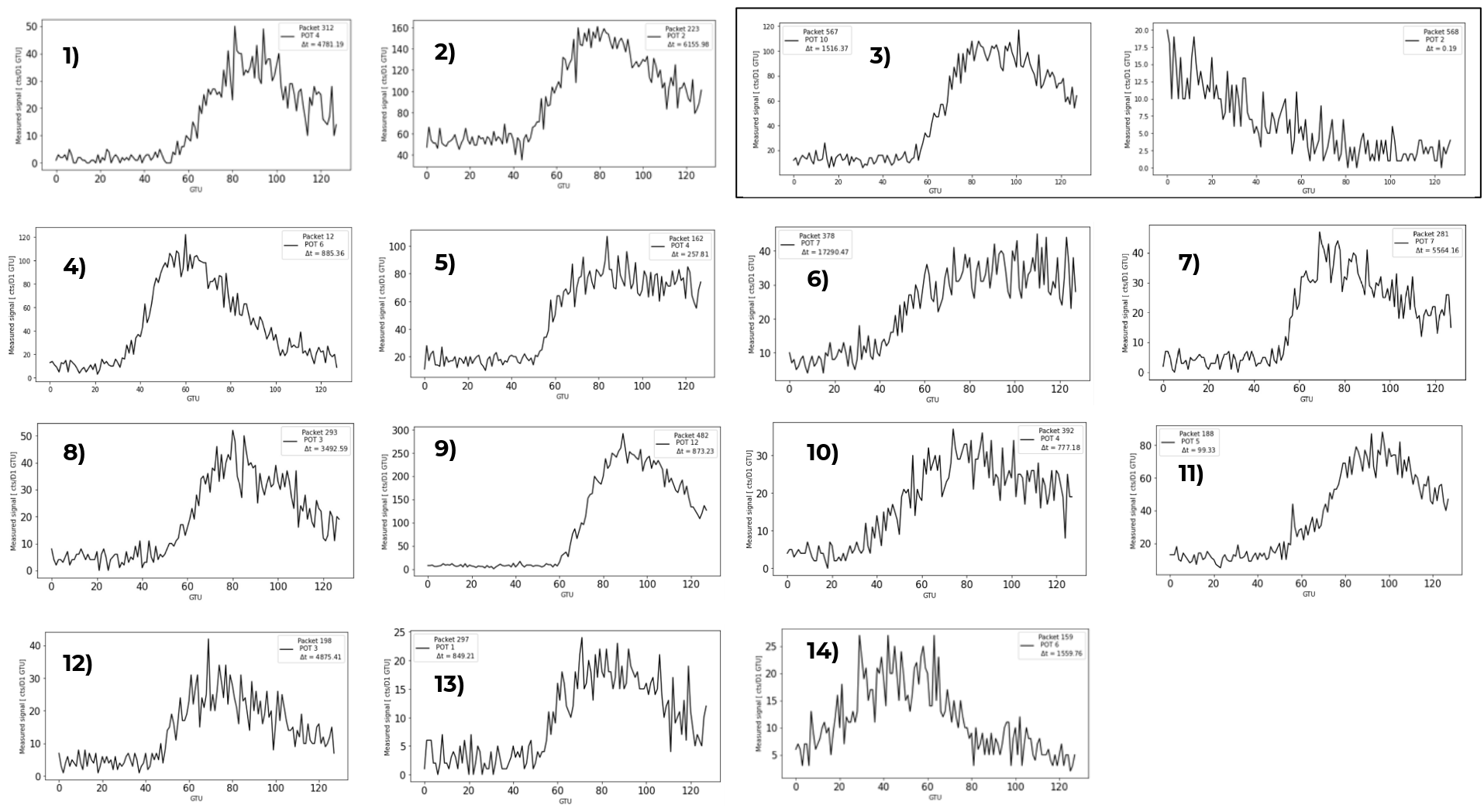}
\caption{Ligtcurves of the 14 SLTs. For events 1, 2, 6, 8, 12, and 13 an atmospheric event has been detected from the exact same position within a few ms after the SLT events.}
\label{Fig:EAS_like_events__lightcurves}
\end{figure}

Short Light Transients (SLTs) are non anthropogenic flashing signals lasting no more than 200~$\mu$s (therefore observed in the D1 dataset). The analysis of a subset of the dataset identified 14  SLT candidates whose lightcurves are shown in Fig.~\ref{Fig:EAS_like_events__lightcurves}. Tipically, these events present a lightcurve with a relatively long signal lasting more than 200~$\mu$s. None of those 14 events present an apparent movement in the focal plane but appear as a stationary light confined in a small cluster of pixels.

The origin of those fast flashing lights is still under study, but it is reasonable to assume that at least some of them are associated with thunderstorm activity in the atmosphere. For 6 SLT events, in fact, Mini-EUSO detected an atmospheric event in the exact same position shortly after the SLT.  The time interval between the two events can vary from less than 1~ms, between 1 and 4~ms or up to 200~ms. We are currently working on the identification of those 6 atmospheric events, that appear to belong to the class of Transient Luminous Events (TLEs) rather than to be more common thunder strikes. We are also investigating any possible correlation between any SLT event with Terrestrial Gamma-ray Flashes (TGFs), which are known to be linked to thunderstorm activities~\cite{ASIM_TGFs}. 

\begin{figure}[hbtp]
\centering
\includegraphics[width=.98\textwidth]{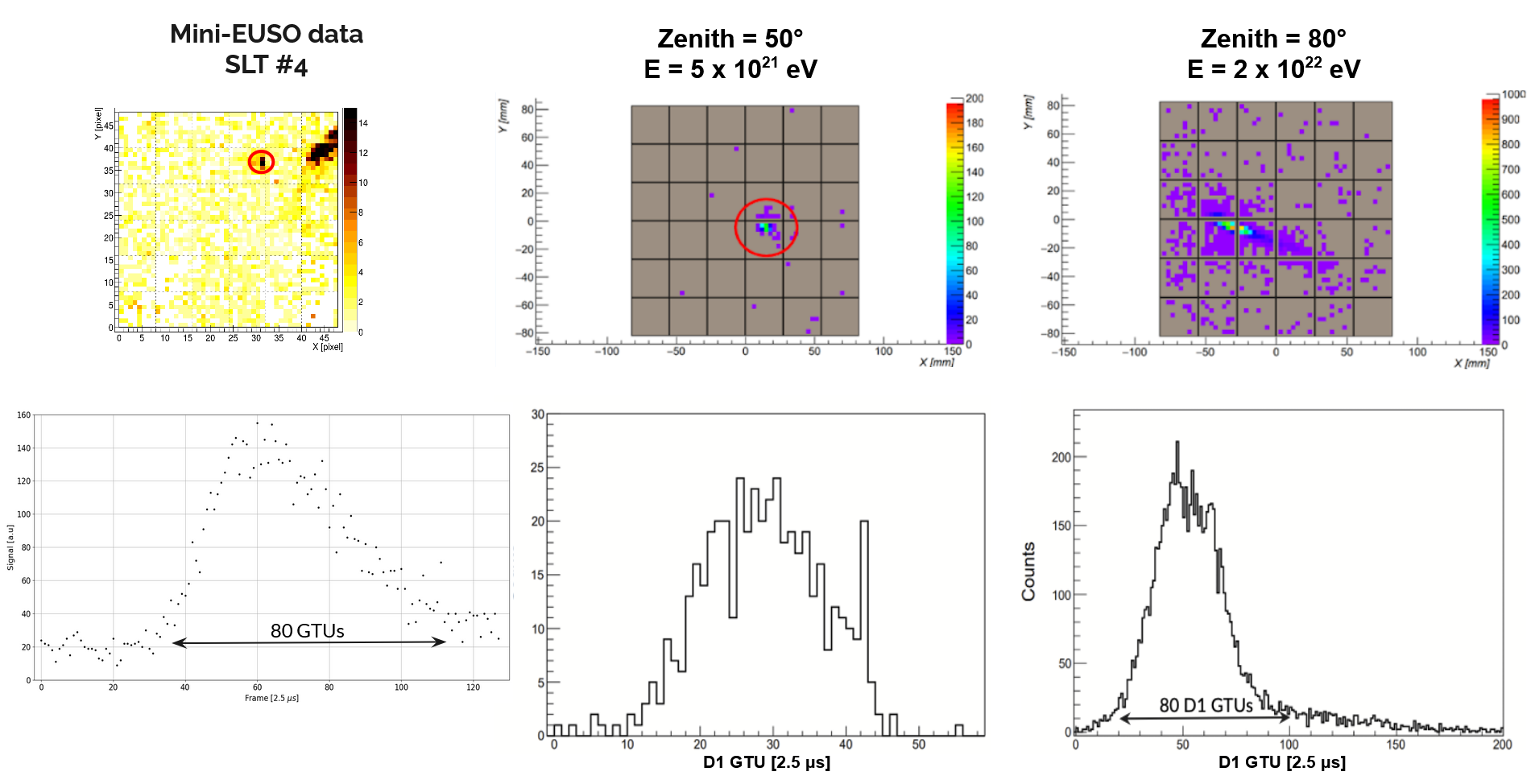} 
\caption{\textbf{Left:} SLT event \#4, the shortest in time. \textbf{Central and right:} ESAF~\cite{ESAF} simulation of protons with different energies and zenithal angles. The signal of a 50$^\circ$ shower is clearly visible and lasts for $\sim$30~D1~GTUs ($\sim$75~$\mu$s). The signal is truncated when the shower reaches the ground. At 80$^\circ$ zenith angle the lightcurve is not truncated (as the shower develops before reaching the ground)  and the signal lasts for $\sim$80~D1~GTUs ($\sim$200~$\mu$s). }
\label{fig:SLT_vs_UHECR}
\end{figure}

The cosmic origin can be excluded with certainty for all the events, also the ones without a follow-up TLE, by a comparison of the focal plane footprint and the lightcurve duration with the cosmic ray simulations shown in Fig.~\ref{fig:SLT_vs_UHECR}. All the 14 events classified as SLTs appear in the focal plane as a small cluster of bright pixels, not too different from the footprint of the 50$^\circ$ zenith angle simulated cosmic ray in Fig.~\ref{fig:SLT_vs_UHECR}. Their lightcurve is however much longer, with a minimum duration of $\sim$80~GTUs ($\sim$200~$\mu$s) for event number 4 and 14, not too distant from the lightcurve of the 80$^\circ$ zenith angle event. For all the events it is therefore not possible to reproduce at the same time the lightcurve shape and duration and the focal plane footprint with a UHECR simulation. This argument by itself is enough to exclude their UHECR origin.

This is an important achievement of Mini-EUSO in view of a future space-based UHECR detector. It demonstrates at the same time that it is possible to trigger from space on something that resembles in many ways the signal of a UHECR, while showing that we are able to discriminate between real cosmic ray signals and other sources.

\section*{Acknowledgements}
This work was supported by the Italian Space Agency (ASI) through the agreement between
ASI and University of Roma Tor Vergata n. 2020-26-HH.0, by the French space agency CNES,
and by the National Science centre in Poland grant 2020/37/B/ST9/01821. This research has been
supported by the Russian State Space Corporation Roscosmos.

\bibliography{my-bib-database}

 \newpage
{\Large\bf Full Authors list: The JEM-EUSO Collaboration}
%{\scriptsize (author-list as of August 20th, 2025 with reorganized affiliations)} \hspace{0.6cm}
%{\scriptsize (version  \today{} \currenttime{})}
%\vspace*{0.5cm}
%contact: zbigniew.plebaniak@roma2.infn.it, marco.ricci@lnf.infn.it

\begin{sloppypar}
{\small \noindent
M.~Abdullahi$^{ep,er}$              % Italy
M.~Abrate$^{ek,el}$,                % Italy
J.H.~Adams Jr.$^{ld}$,              % USA 
D.~Allard$^{cb}$,                   % France
P.~Alldredge$^{ld}$,                % USA
R.~Aloisio$^{ep,er}$,               % Italy
R.~Ammendola$^{ei}$,                % Italy
A.~Anastasio$^{ef}$,                % Italy %%
L.~Anchordoqui$^{le}$,              % USA
V.~Andreoli$^{ek,el}$
% ...
}
\end{sloppypar}
\vspace*{.3cm}

%%\newpage
{ \footnotesize
\noindent
%
% Czech Republic - 2 intitutions
%Czech Republic\\
% ...
$^{cb}$ Universit\'e de Paris, AstroParticule et Cosmologie (CNRS), Paris, France\\
% ...
$^{ef}$ Istituto Nazionale di Fisica Nucleare (INFN), Sezione di Napoli, Naples, Italy\\
% ...
$^{ei}$ Istituto Nazionale di Fisica Nucleare (INFN), Sezione di Roma Tor Vergata, Rome, Italy\\
% ...
$^{ek}$ Istituto Nazionale di Fisica Nucleare (INFN), Sezione di Torino, Turin, Italy\\
$^{el}$ Universit\`a di Torino, Dipartimento di Fisica, Turin, Italy\\
% ...
$^{ep}$ Gran Sasso Science Institute (GSSI), L’Aquila, Italy\\
$^{er}$ Istituto Nazionale di Fisica Nucleare (INFN), Laboratori Nazionali del Gran Sasso, Assergi, Italy\\
% ...
$^{ld}$ University of Alabama in Huntsville, Huntsville, AL, USA\\
$^{le}$ City University of New York (CUNY), Lehman College, Bronx, NY, USA\\
% ...
}

\end{document}